\documentclass[superscriptaddress,twocolumn,showpacs,showkeys,amsmath,prb,final]{revtex4}

\usepackage{graphicx}
\usepackage{dcolumn}
\usepackage[sort&compress]{natbib}
\usepackage{subfigure}
\usepackage{ifpdf}
\usepackage{bm}
\usepackage{hyperref}

\ifpdf
\usepackage{hyperref}
\hypersetup{%
pdfauthor={Mauro M. Doria, Antonio R. de C. Romaguera, and S.
Salem-Sugui Jr},
pdftitle={Vanishing of the upper critical field in
Bi$_2$Sr$_2$CaCu$_2$O$_{8+\delta}$ from Landau-Ott scaling},
pdfsubject={Superconductivity},
pdfkeywords={Department of Physics , Universiteit Antwerpen
 Groenenborgerlaan 171 / U 208 B-2020 Antwerpen},
baseurl={http://www.if.ufrj.br/~mmd} }
\fi

\begin{document}
\title{Vanishing of the upper critical field in Bi$_2$Sr$_2$CaCu$_2$O$_{8+\delta}$ from Landau-Ott scaling}
\author{Mauro M. Doria}%
\email{mmd@if.ufrj.br}\homepage{http://www.if.ufrj.br/~mmd}%
\affiliation{Instituto de F\'{\i}sica, Universidade Federal do Rio
de Janeiro, C. P. 68528, 21941-972, Rio de Janeiro, Brazil}%
\affiliation{Departement Fysica, Universiteit Antwerpen,
Groenenborgerlaan 171, B-2020
Antwerpen, Belgium}%
\author{Antonio R. de C. Romaguera}%
\affiliation{Instituto de F\'{\i}sica, Universidade Federal do Rio
de Janeiro, C. P. 68528, 21941-972, Rio de Janeiro, Brazil}%
\affiliation{Departement Fysica, Universiteit Antwerpen,
Groenenborgerlaan 171, B-2020 Antwerpen, Belgium}
\author{S. Salem-Sugui Jr}%
\affiliation{Instituto de F\'{\i}sica, Universidade Federal do Rio
de Janeiro, C. P. 68528, 21941-972, Rio de Janeiro, Brazil}%
\date{\today}

\begin{abstract}
We apply Landau-Ott scaling to the reversible magnetization data of
Bi$_2$Sr$_2$CaCu$_2$O$_{8+\delta}$ published by Y. Wang et al.
[\emph{Phys. Rev. Lett. \textbf{95} 247002 (2005)}] and find that
the extrapolation of the Landau-Ott upper critical field line
vanishes at a critical temperature parameter, $T^* _c$, a few
degrees above the zero resistivity critical temperature, $T_c$. Only
isothermal curves below and near to $T_c$ were used to determine
this transition temperature. This temperature is associated to the
disappearance of the mixed state instead of a complete suppression
of superconductivity in the sample.
\end{abstract}
\preprint{APS/123-QED}
\maketitle

There are conflicting views about the nature of the upper critical
field $H_{c2}(T)$ in the present literature, possibly because this
concept involves multiple distinct phenomena. The traditional
Abrikosov's~\cite{abrikoJETP57} view is of densely packed vortices
with nearly touching cores that make the normal state percolate
inside the superconducting state. The collapse happens at a well
defined temperature $T_c$ because the coherence length, $\xi(T)$,
that sets the vortex core area, diverges at $T_c$ making the upper
critical field, $H_{c2}(T)=\Phi_0/2\pi\xi(T)^2$, vanish there
($H_{c2}(T\rightarrow T_c)\rightarrow 0$). Recently this view was
challenged by Y. Wang et al.~\cite{ongPRL05,ongPRB06} who proposed a
quite distinct scenario for Bi$_2$Sr$_2$CaCu$_2$O$_{8+\delta}$ (Bi
2212). Their $T_c$ just sets the loss of phase coherence but not of
the diamagnetic superconducting signal. Therefore they suggested
that Cooper pairs still exist above $T_c$, and so do the vortices.
Consequently the field $H_{c2}(T)$ does not vanish at $T_c$ and in
fact can be quite large there. Their view has grown out of Nernst
effect~\cite{ongPRL05} and sensitive torque
magnetometry~\cite{ongPRB06} experiments. The latter remarkably
contain isothermal magnetization curves below and also {\it above}
$T_c$. They report on two new temperatures above $T_c$, the highest
one, $T^*$, associated with local correlations affecting spin
degrees of freedom and the lowest one, $T_{onset}$, with the onset
of vorticity and supercurrents. The vanishing of the upper critical
field takes place at a much higher temperature, where the Nernst
signal extrapolates to zero. In practical terms the upper critical
field becomes an inherently unmeasurable quantity for the high-$T_c$
materials~\cite{ongPRB06} in this scenario.

In this paper we apply a scaling method developed by I. L. Landau and H. R. Ott~\cite{landottPRB02} to the
reversible magnetization data obtained from the torque magnetometry measurements of Y. Wang et al. and report a
new temperature $T^*_c$, above $T_c$, which does not coincide with any of the above temperatures and in fact is
much lower than $T_{onset}$.
\begin{figure}[b]
\centering
\includegraphics[width=\linewidth]{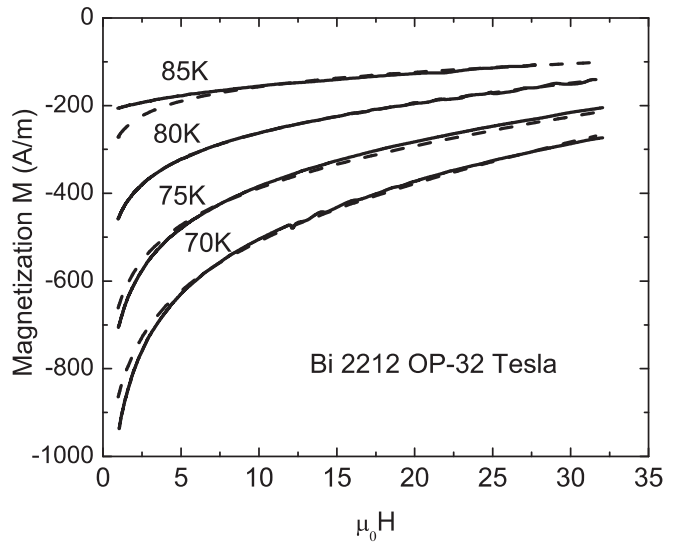}
\caption{Isothermal magnetization versus applied field lines are shown here. The solid lines are retrieved from
Ref.~\onlinecite{ongPRL05} for the (OP) Bi2212 compound, and correspond to the temperatures of 70, 75, 80, and
85 K. The dashed lines are polynomial fits obtained through Eq.(\ref{eq3}), and are referred as $M_{eff}$ curves
in the text.} \label{figa}
\end{figure}
The upper critical field and the Nernst effect are related through
the transport entropy per unit length of the vortex line. In fact
the Nernst coefficient is just the product of the transport entropy
per unit length of the vortex line and the resistivity. The
Caroli-Maki-Hu relation~\cite{makiJLTP69,makiPTP69,huPRB76,huPRB76b}
provides the way to connect the transport entropy per unit length of
the vortex line to the reversible magnetization $M(H,T)$, which in
turn leads to the upper critical field. This last connection can be
achieved, for instance, using the celebrated Abrikosov's
expression~\cite{abrikoJETP57,serinRMP64}:
\begin{equation}
M(H,T)=\frac{H_{c2}(T)-H}{\beta_A(2\kappa^2-1)}, \label{eq1}
\end{equation}
where $\beta_A$ is a constant that depends on the vortex arrangement
and $\kappa$ is the Ginzburg-Landau parameter, an intrinsic property
of the superconductor. It turns out that this expression is also the
starting point for Landau and Ott who proposed a traditional view of
the upper critical field and applied it successfully to several of
the high-$T_c$
materials~\cite{landottPRB02,landottPRB03,landottPC03,landottPC04,landottPRB05,landottPRB05b},
including Bi 2212. Their proposal renders a scaling method that can
be directly sought in the transport entropy per unit length of the
vortex line obtained from the Nernst effect, but this is not done
here. Remarkably this scaling procedure retrieves a nearly linear
borderline in the H vs. T diagram very similar to the original
Abrikosov proposal described by Eq.(\ref{eq1}). This is quite a
surprising fact, considering that these boundary lines for the
high-$T_c$ materials, including Bi-2212, usually display an upward
(positive) curvature, such as for the irreversibility
line~\cite{bedmullerPRL87} and also for the melting transition
line~\cite{delacruzPRL94,zeldovN95}.
\begin{figure}[b]
\centering
\includegraphics[width=\linewidth]{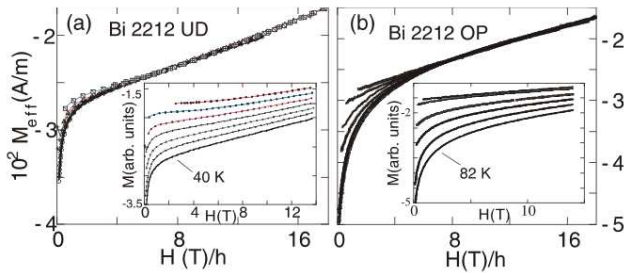}
\caption{Isothermal curves (UD: $T$=40, 41, 42, 43, 44, 45 and 46 K;
OP: 82, 83, 84, 85, 86 and 86.3 K) are scaled according to the
Landau-Ott scaling.} \label{fig1}
\end{figure}
The Landau-Ott approach relies on the very basic assumption that the
magnetic susceptibility $\chi(h) \equiv M(H,T)/H$ is a sole function
of the reduced field $h=H/H_{c2}(T)$, such that all its temperature
dependence is contained in the upper critical field. Their proposal
is inspired by Eq.(\ref{eq1}), which does satisfy this condition in
case that $\kappa$ is a temperature independent parameter. From this
assumption the scaling relation connecting magnetization values at
two different temperatures, $T_0$ and $T$ may be derived as follows:
\begin{equation}
M(H,T_{0})=M(h_{c2}H,T)/h_{c2}, \label{eq2}
\end{equation}
$h_{c2}=H_{c2}(T)/H_{c2}(T_{0})$. This relation implies that all isothermal reversible magnetization curves
collapse into a single curve by a judicious choice of the parameter $h_{c2}(T)$. The collected set of scaling
parameters $h_{c2}(T)$, once plotted versus $T$, leads to the curve $H_{c2}(T)$ once $H_{c2}(T_{0})$ is
explicitly known. In this way Landau and Ott retrieved their $H_{c2}(T)$ curve from the background free
reversible magnetization of many high-$T_c$. A direct consequence of their method is the existence of a
temperature parameter, $T^*_c$, where the upper critical field extrapolates to zero: $H_{c2}(T^*_c)=0$. This
temperature has been found to coincide with $T_c$ for the high-$T_c$
materials~\cite{landottPRB02,landottPRB03,landottPC03,landottPC04,landottPRB05,landottPRB05b}, a fact that has
been invoked by Landau and Ott as indicative of the correctness of their method. Recently their analysis was
applied to the low-$T_c$ materials~\cite{doriaPRB06,landPC07} and there these two temperatures were also found
to coincide.

We report here that the $H_{c2}(T)$ curve, as obtained from the torque magnetometry data of Y. Wang et al. for
Bi2212, does not vanish at $T_c$ according to the Landau-Ott scaling. Y. Wang et al. considered two compounds,
underdoped (UD) and optimally doped (OP) Bi 2212, with $T_c$ equal to 50 K and 87.5 K, respectively. Two kinds
of field sweeps were used to obtain their isothermal curves. The first one takes a field range up to 14 Tesla
for both the UD and OP compounds, and the second a range of 32 Tesla, but in this last case measurements were
only taken for the OP compound. We applied the Landau-Ott scaling to these data sets and found the striking
result that $T^*_c$ is equal to 57 K for the UD compound and 93 K for the OP compound, significantly higher than
the corresponding $T_c$ values, even considering the maximum error bar of 0.8 K in our calculations. To
determine the temperature $T^*_c$ we have only considered Y. Wang et al.'s isothermal magnetization curves that
fall under two conditions: (1)below and (2) close to $T_c$. This is the temperature range that Landau-Ott
scaling must hold although it was found to hold relatively much below $T_c$ for some low-temperature
compounds~\cite{doriaPRB06}.

Fig.~\ref{figa} shows the fitting of four isothermal magnetization
curves to the polynomial
\begin{equation}
M_{eff}(H) = h_{c2}{ \sum_{i=0}^n A_i [ln(H/h_{c2})]^i + c_0 H}. \label{eq3}
\end{equation}
Y. Wang et al. reported that their fully reversible magnetization
data has the paramagnetic background carefully removed. We notice
that a parameter $c_0$ still had to be included here. This parameter
is part of the Landau-Ott prescription for the removal of a residual
background field. Fig.~\ref{figa} shows (solid lines) the four
closest curves to $T_c$ obtained from Fig.4(a) of
Ref.~\onlinecite{ongPRL05} for the OP compound that belong to the 32
Tesla data set: 70, 75, 80 and 85 K. The 80 K data curve is taken as
the reference curve $(h_{c2}=1)$. For instance, we achieved a fairly
good description of it through a fourth order polynomial ($n=4$)
with coefficients $(A_4, A_3, A_2, A_1, A_0)$ equal to $(+0.6438,
-2.2732, +2.9544, +81.0093, -455.1729)$. The remaining three
isothermal curves are also fitted by this polynomial with the
$(h_{c2}, c_0)$ parameters equal to  $(1.73, -1.55)$, $(1.37, -0.7)$
and $(0.64, 2.5)$ for the 70, 75, and 85 K curves, respectively. The
average mean square deviation from this fit is of the order of $3\%$
for these three curves and of the order of $0.07\%$ for the 80 K
curve.
\begin{figure}[b]
\centering
\includegraphics[width=\linewidth]{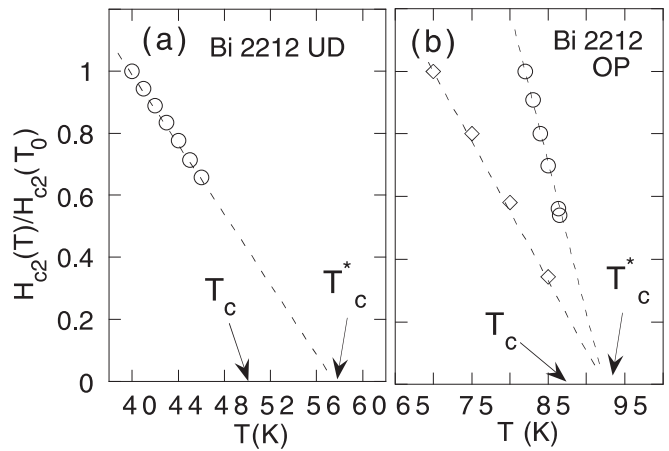}
\caption{$H_{c2}(T)$  obtained through the Landau-Ott scaling
normalized by the $T_0$=40 K (OD-14 \textrm{Tesla}), 70 K (OP-32
\textrm{Tesla}) and 85 K (OP-14 \textrm{Tesla}) curves. The linear
extrapolation to zero field defines the temperature $T^*_c$. }
\label{fig2}
\end{figure}
The same kind of polynomial analysis was applied for the 14 Tesla data sets, obtained from Figs. 2(a) and 2(b)
of Ref.~\onlinecite{ongPRL05} for the UD and OP compounds, respectively. The major plots of Fig.~\ref{fig1} show
the collapsed curves after the Landau-Ott scaling. The original isothermal magnetization curves are shown in the
insets. Notice that firstly a residual background must be removed to obtain $M_{eff}$, as previously explained.
Fig.~\ref{fig2} shows the collected scaling parameters $h_{c2}(T)$ for the selected set of temperatures.
Surprisingly their linear fit extrapolates away from $T_c$, revealing the existence of the new temperature
parameter $T^*_c$. Notice that this analysis was done for the OP compound using both the 14 Tesla and the 32
Tesla  data sets and both render virtually the same $T^*_c$, with less than 0.8 K difference. We stress that the
present results are invariant under the choice of the reference isothermal curve. To check this we have taken in
Fig.~\ref{fig2} the 70 K curve as the reference isothermal curve, thus differently from Fig.~\ref{figa}, which
takes the 80 K curve instead. Again we have obtained the same $T^*_c$ under the same precision window.

Fig.~\ref{figa} shows that the polynomial fits break down at low
field. These fits are good within a window of nearly 25 Tesla that
does not include the approximately initial 5 Tesla range. In this
low field range the fits overestimate the diamagnetism well below
$T_c$ and underestimate it close to $T_c$. So it is conceivable that
$H_{c2}$ data points of Fig.~\ref{fig2} can be affected by this
polynomial fit break down. Besides, the $H_{c2}$ points, lower than
those reported in Fig.~\ref{fig2}, could make this line turn down
and extrapolate to the observed $T_c$ value or, at least, to a
$T^*_c$ lower than the values indicated in Fig.~\ref{fig2}.

In support for the existence of this new critical temperature
$T^*_c$ we notice that long ago Huebener and
co-workers~\cite{huebPRB93,huebPRB94,huebSST95} had to introduce a
new temperature in their best fit analysis of the reversible
magnetization obtained from the Nernst effect. In other words, the
YBa$_2$Cu$_3$O$_{7-\delta}$ data best fit to Eq.(\ref{eq1}) yields
an upper critical field that extrapolates to zero away from $T_c$.
They reported a zero resistivity transition temperature equal to
93.0 K (see Table I of Ref.~\onlinecite{huebPRB94}) which does not
coincide with the higher temperature of 93.8 K found by
extrapolation of their upper critical field line to zero (see Fig. 8
of Ref.~\onlinecite{huebPRB94}).

In conclusion we find here  that the Landau-Ott scaling applies to Y. Wang et al.'s data and removes the
extremely large upper critical field values near to $T_c$ because now $H_{c2}(T)$ vanishes at $T^*_c$, a new
parameter not considered in their analysis. Y. Wang et al. fitted Bi 2212 and also NbSe$_2$ magnetic torque data
to $M \sim - [H_{c2}(T)-H]$ (Eq.(\ref{eq1})) to show that the high-$T_c$ materials have unusual behavior as
compared to the low-$T_c$ materials. They found that Abrikosov's picture holds for NbSe$_2$, since $H_{c2}(T)$
vanishes near $T_c$, but not for Bi 2212, where it is extremely large: $H_{c2}$(86 K) = 90 Tesla! According to
the Landau-Ott view~\cite{landottPRB02} the $H_{c2}(T)$ curves of Fig.~\ref{fig2} set the disappearance of the
mixed state rather than to a complete suppression of superconductivity in the sample. Thus the present view of
$H_{c2}(T)$ is not inconsistent with incoherent superconductivity above $T_c$, whose onset and disappearance
must be referred by names other than $H_{c2}(T)$.

We thank CNPq (Brazil), FAPERJ (Brazil) and the Instituto do
Mil\^enio de Nanotecnologia (Brazil) for financial support. M.M.
Doria acknowledges support from BOF/UA (Belgium).
\\

\begin{thebibliography}{99}
\bibitem{abrikoJETP57} A. A. Abrikosov, Soviet Physics JETP {\bf 5}, 1174 (1957).
\bibitem{ongPRB06} Yayu Wang, Lu Li and N. P. Ong., Phys. Rev. B {\bf 73}, 024510 (2006).
\bibitem{ongPRL05} Yayu Wang, Lu Li, M. J. Naughton, G. D. Gu, S. Uchida and N. P. Ong, Phys. Rev. Lett. {\bf 95}, 247002 (2005).
\bibitem{landottPRB02} I. L. Landau and H. R. Ott, Phys. Rev. B {\bf 66}, 144506 (2002).
\bibitem{makiJLTP69} Kazumi Maki, J. of Low Temp. Phys. {\bf 1}, 45 (1969).
\bibitem{makiPTP69} Kazumi Maki, Progr. of Theo. Phys. {\bf 41}, 902 (1969).
\bibitem{huPRB76}Chia-Ren Hu, Phys. Rev. B {\bf 13}, 4780 (1976).
\bibitem{huPRB76b}Chia-Ren Hu, Phys. Rev. B {\bf 14}, 4834 (1976).
\bibitem{serinRMP64} T. Kinsel, E. A. Lynton and B. Serin, Rev. Mod. Phys. {\bf 36}, 105 (1964).
\bibitem{landottPRB03} I. L. Landau and H. R. Ott, Phys. Rev. B {\bf 67}, 0922505 (2003).
\bibitem{landottPC03}I. L. Landau and H. R. Ott, Physica C {\bf 398}, 73 (2003).
\bibitem{landottPC04}I. L. Landau and H. R. Ott, Physica C {\bf 411}, 83 (2004).
\bibitem{landottPRB05} I. L. Landau and H. R. Ott, Phys. Rev. B {\bf 71}, 012511 (2005).
\bibitem{landottPRB05b} I. L. Landau and H. R. Ott, Phys. Rev. B {\bf 72}, 176502 (2005).
\bibitem{bedmullerPRL87} K. A. M\"uller, M. Takashige  and J. G. Bednorz, Phys. Rev. Lett., {\bf 58}, 1143 (1987).
\bibitem{delacruzPRL94} H. Pastoriza, M. F. Goffman, A. Arrib\'ere and F. de la Cruz, Phys. Rev. Lett., {\bf 72}, 2951 (1994).
\bibitem{zeldovN95} E. Zeldov, D. Majer, M. Konczykowski, V. B. Geshkenbein, V. M. Vinokur,  and H. Shtrikman, Nature {\bf 375} 373 (1995).
\bibitem{doriaPRB06} Mauro M. Doria, S. Salem-Sugui, Jr., P. Badica and K. Togano {\bf 73}, 184524 (2006).
\bibitem{landPC07}I.L. Landau, R. Khasanov, K. Togano  and H. Keller, Physica C {\bf 451}, 134 (2007).
\bibitem{huebPRB93}  H.-C. Ri,  F. Kober, A. Beck, L. Alff, R. Gross, and R. P. Huebener, Phys. Rev. B {\bf 47}, 12312 (1993).
\bibitem{huebPRB94} H.-C. Ri,R. Gross, F. Gollnik, A. Beck, R. P. Huebener, P. Wagner and H. Adrian, Phys. Rev. B {\bf 50} 3312, (1994).
\bibitem{huebSST95} R P Huebener, Supercond. Sci. and Technol.{\bf 8}, 189 (1995).
\end{thebibliography}

\end{document}